\documentclass{aastex631}

\usepackage{bm}
\usepackage{xcolor}
\usepackage{multibib}
\newcites{main}{References}
\newcites{appendix}{References (Appendix)}
\usepackage{subfigure}
\usepackage{soul}
\usepackage{tcolorbox}
\usepackage{amsmath}
\usepackage{verbatim}
\newcommand{\beq}{\begin{equation}}
\newcommand{\eeq}{\end{equation}}
\newcommand{\bseq}{\begin{subequations}}
\newcommand{\eseq}{\end{subequations}}
\newcommand{\bary}{\begin{eqnarray}}
\newcommand{\eary}{\end{eqnarray}}
\newcommand{\bwt}{\begin{widetext}}
\newcommand{\ewt}{\end{widetext}}

\begin{document}

\title {Photohadronic interpretation of the different incarnation of 1ES 2344+514}

\author{Sarira Sahu}
\email{sarira@nucleares.unam.mx}
\affiliation{Instituto de Ciencias Nucleares, Universidad Nacional Aut\'onoma de M\'exico, \\
Circuito Exterior, C.U., A. Postal 70-543, 04510 México DF, México
}

\author{Isabel Abigail Valadez Polanco}
\email{abivaladez@gmail.com}
\affiliation{Facultad de Ingeniería, Universidad Aut\'onoma de Yucat\'an, \\ 
Industrias No Contaminantes S/N, Sin Nombre de Col 27, M\'erida, Yucat\'an, M\'exico
}
\author{Subhash Rajpoot}
\email{Subhash.Rajpoot@csulb.edu}
\affiliation{Department of Physics and Astronomy, California State University,\\ 
1250 Bellflower Boulevard, Long Beach, CA 90840, USA
}

\begin{abstract}
Since its discovery in 1995, the high-energy peaked blazar 1ES 2344+514 has undergone several episodes of GeV-TeV flaring and has been observed in the multiwavelength by several other telescopes. The observed X-ray spectrum of 1996 and the flaring event of 2016 establish that 1ES 2344+514 has a temporary EHBL-like behavior.
Such behavior has also been observed in several nearby high-energy peaked blazars. We use the photohadronic model to account for the GeV-TeV flaring observed events of 1995 and 2007. Also, a recently proposed two-zone photohadronic model, which is successful in explaining the multi-TeV flaring events of many transient EHBL-like source, is employed to explain the GeV-TeV flaring spectra of MJD 57611 and MJD 57612. We find that the zone-2 parameters of the  two-zone photohadronic model play a central role in explaining these spectra. Probably this is an indication of a new type of transient EHBL-like source. We find that our fits to the observed spectra are comparable or better than the other leptonic and hadronic models employed in the literature to address the same issue.
\end{abstract}


\section{Introduction}
Blazars are the primary sources of extragalactic $\gamma$-rays. Their spectral energy distribution (SED) consists of two non-thermal peaks in the $\nu$-$F_\nu$ plane \citep{Dermer:1993cz}. The low energy peak in the infrared to the X-ray frequency region is due to the synchrotron photons produced from a population of relativistic electrons in the jet magnetic field. It is believed that the high energy peak between the X-ray and the $\gamma$-ray energies is produced either from the Synchrotron Self-Compton (SSC) scattering of high-energy electrons and the self-produced synchrotron photons in the jet \citep{Maraschi:1992iz,Gao:2012sq} or due to the up-scattering of low energy external seed photons and the ultra-relativistic electrons in the jet. The sources of the seed photons could be either the accretion disk or the broad line regions or the dusty torus \citep{Blazejowski:2000ck,Sikora:1994zb,Dermer:1993cz}. Recently, different radiation mechanisms related to the leptonic and the hadronic processes were reviewed in the context of AGN jets by \cite{Cerruti:2020lfj}. Blazars are classified according to the position of the synchrotron peak frequency as,  
low-energy peaked blazars (LBLs, $\nu_{peak} < 10^{14}\, Hz$), intermediate-energy peaked blazars (IBLs, $10^{14}\, Hz < \nu_{peak}  < 10^{15}\, Hz$), high energy-peaked blazars (HBLs, $10^{15}\, Hz <  \nu_{peak} \, < 10^{17}\, Hz$) \citep{Abdo:2009iq} and extreme energy peaked blazar (EHBL, $ >10^{17}$ Hz) \citep{Costamante:2001pu}.

The EHBLs have unusually high synchrotron peak frequency and also the SSC peak shifts towards the higher energy regime of $\gamma$-rays. Therefore, the shift of the synchrotron peak and the SSC peak in the multiwavelength (MW) SED generally translate to a particularly hard X-ray and a very high energy (VHE, $>\, 100$ GeV) $\gamma$-ray spectra.
Low luminosity and limited variability are the characteristic features of EHBLs which make them difficult to detect \citep{Costamante:2017xqg,Acciari:2019jzj}. Thus, the
population of the observed EHBLs are much less than the observed HBLs 
\citep{Aharonian:2007nq,2007A&A...473L..25A,2007A&A...475L...9A}.
Different temporal bahavior has been observed among the EHBLs. For example, 1ES 0229+200, 1ES 0347-232, RGB J0710+591 and 1ES 1101-232 \citep{Aharonian:2007nq,Costamante:2017xqg} are the archetypal EHBLs which constantly exhibit extreme properties. However, some well known and well studied nearby HBLs
such as Markarian 421 (Mrk 421), Markarian 501 (Mrk 501) and 1ES 1959+650 belong to the EHBL family as temporary members \citep{Ahnen:2018mtr,Foffano:2019itc}.
In 1996 and 2016 another comparatively nearby HBL 1ES 2344+514 had also shown a temporary EHBL-like behavior and a hard TeV spectra \citep{2020MNRAS.496.3912M}. Particularly, during the flaring of 2016, both the synchrotron and the SSC peaks were pushed towards higher energies
\citep{Ahnen:2018mtr,2020A&A...638A..14M,Aleksi__2015}.
In March 2010, Mrk 421 was very active and various telescopes made observations in the MW range for 13 consecutive days \citep{Aleksi__2015}. The synchrotron peak position was above $10^{17}$ Hz and the second peak was also at higher energies during this observation period, thus exhibiting the signature for extreme HBL-like behavior. Mrk 501 exhibited EHBL-like behavior during the VHE flaring of May - July 2005 \citep{Albert:2007zd} and June 2012 \citep{Ahnen:2018mtr}. Similarly, during a MW campaign of 1ES 1959+650 between April 29 and November 21 of 2016, the MAGIC telescopes observed multi-TeV flaring events for the nights of 13th, 14th of June and 1st of July 2016 when the position of 
the synchrotron peak was in the EHBL range, and again, exhibiting 
extreme HBL-like behavior \citep{2020A&A...638A..14M}. Several other authors have also studied the MW SED of 1ES 1959+650 \citep{2021MNRAS.504.5485S,2021ApJ...918...67C}.

The accumulating evidence from a large number of MW observations so far suggests that the one-zone SSC model is too simple to account for the SEDs of these objects. In the model, the emission from optical to VHE $\gamma$-rays takes place in a single spherical blob. Moreover, in order to explain the observed spectra, large values of the minimum electron Lorentz factor, the bulk Lorentz factor \citep{Ahnen:2018mtr,2020A&A...638A..14M}, and unrealistically low values of the magnetic field are used. In order to avoid the above shortcomings, alternative models have been proposed, such as the two-zone leptonic model, the inverse Compton scattering of relativistic electrons with the cosmic microwave background, the spine-layer structured jet model and different other hadronic models \citep{2020MNRAS.496.3912M,2020A&A...640A.132M}.

Previously, the VHE spectra of many HBLs and EHBLs were successfully explained using the photohadronic model \citep{Sahu:2019kfd,Sahu:2019scf}. The photohadronic model is based on the assumption that during the VHE flaring event, the Fermi accelerated high energy protons in the inner jet region interact with the low energy tail region of the  SSC seed photons  
through the dominant $\Delta$-resonance channel $p\gamma\rightarrow\Delta^+$ with the multi-pion channels suppressed. Thus, due to this process, a change in the spectral behavior of the SSC photons is expected in the VHE spectrum. The $\Delta$-resonance subsequently decays to $\gamma$-rays via neutral pion production in the jet environment. These  $\gamma$-rays constitute the observed VHE spectrum. 
However, given the kinematical conditons and the density of the background seed photons in the internal jet, the photohadronic process works well only for $E\gamma \gtrsim 100$ GeV.  Below this energy the leptonic processes have the most important contribution to the multiwavelength SED. Here, we use the photohadronic process to interpret the VHE spectrum \citep{Sahu:2019lwj}.
In several previous studies, we have shown that the flux in the low energy tail region of the SSC spectrum can be expressed as a power-law which is taken to be as $\Phi_{SSC}\propto E_{\gamma}^{-\beta}$ where $E_{\gamma}$ is the observed  photon energy and $\beta$ is the spectral index. The value of $\beta$ lies in the range $0 < \beta \le 1$ to explain the low emission state, the high emission state and the very high emission state of the flaring HBLs \citep{Sahu:2019kfd}. Recently, it was shown that the temporary EHBL-like behavior of the flaring events of Mrk 421, Mrk 501 and 1ES 1959+650 are difficult to explain by the  photohadronic model \citep{Sahu:2020tko,Sahu:2020kce,Sahu:2021wue} in its simplest form. This happens because $\Phi_{SSC}$ fails to behave as a single power-law for the temporary EHBL. During this temporary EHBL-like behavior, the synchrotron peak and the SSC peak are shifted towards higher energies. Due to this shift, the standard jet parameters, such as the blob size, the bulk Lorentz factor, and the strength of the magnetic field, may differ from the ones used previously in HBL, but the mechanisms to accelerate particles and the emission mechanisms are taken to be the same \citep{Sahu:2021wue}. In this changed environment, we claim that the photohadronic process should still be the dominant process. In order to explain the temporary EHBL-like behavior of the VHE flaring events of the above objects, the photohadronic  model was extended to include two different zones for the SSC flux: the first one with the conventional behavior with $0 < \beta
\le 1$ and the second one with $ 1 < \beta  \le 1.5$ with no other modifications to the proton spectral index in the jet.  
Previously, we have shown that in such scenarios, the photohadronic process is favored over other mechanisms such as the single zone SSC or the proton synchrotron that employ low magnetic field and low cutoff energy for the Fermi accelerated protons.

The HBL 1ES 2344+514 has undergone several VHE flaring episodes since its discovery in 1995. In 1996 the observation in X-rays showed that the synchrotron peak was above $4\times10^{18}$ Hz, a signature of EHBL-like bahavior. Similarly, in August 2016 the ground-based $\gamma$-ray telescope First G-APD Cherenkov Telescope (FACT) detected 1ES 2344+514 in a high emission state  which triggered multiwavelength observations. During this observational period, it was observed that the VHE $\gamma$-ray spectrum and the X-ray spectrum were hard and the frequency shift of the synchrotron peak was towards $\gtrsim 10^{18}$ Hz, a renewal of extreme-EHBL like behavior of 1ES 2344+514. However, during 1995, 2007-2008 and 2014-2015, the VHE flaring events were found to be of the standard HBL types which shows that the shift towards EHBL-like behavior was temporary. Thus, in order to correctly account for the standard HBL and EHBL we employ the one-zone photohadronic model for the VHE flaring events for which the synchrotron peak is below $10^{17}$ Hz and the two-zone photohadronic model to explain the VHE flaring events for which the synchrotron peak is 
$\gtrsim\, 10^{17}$ Hz. We apply the one-zone photohadronic analysis to explain the VHE flaring events of December 1995 and December 2007 of 1ES 2344+514 as the synchrotron peaks were in the HBL limit ($10^{15}-10^{16} $ Hz). We apply the two-zone photohadronic model to the flaring events of a 11th and 12th of August 2016 as observed by the MAGIC telescopes in which the source demonstrated EHBL-like behavior.

\section{Flaring history of 1ES 2344+514}
1ES 2344+514 was first detected 
in the Einstein Slew Survey \citep{1992ApJS...80..257E} at X-ray energies and is classified in the HBL category \citep{1996ApJS..104..251P}. It was the third VHE extragalactic source detected after Mrk 421 and Mrk 501 \citep{Catanese_1998} at a red shift of $ z = 0.044$ \citep{,1996ApJS..104..251P} and a luminosity distance of $d_L\sim 200$ Mpc. For the first time, the VHE $\gamma$-rays from this object were observed during an intense flare on December 20, 1995 by the Whipple 10 m telescope and the observed flux above 350 GeV was almost 60\% of the Crab Nebula flux \citep{Catanese_1998,Schroedter_2005}.
In 1996 a flare in X-ray energy was observed with a large flux variability on timescales of a few hours when the source was at its brightest \citep{2000MNRAS.317..743G}. During this period the lower limit of the synchrotron peak frequency was shifted above $3\times 10^{18}$ Hz, a clear signature of EHBL-like behavior of the source seen for the first time. This extreme behavior of the source  motivated several multiwavelength campaigns to be undertaken on a regular basis as part of the blazar science programme to observe and to model the 1ES 2344+514 broad-band SED using the simultaneous and the quasi-simultaneous data \citep{2007ApJ...662..892A,Godambe_2007,2011ApJ...738..169A,2013A&A...556A..67A}. Subsequently, it was found that the source was mostly in a low emission state during the following years with no signature of any extreme behavior either in the X-ray band or in the $\gamma$-ray \citep{Catanese_1998,2000MNRAS.317..743G,
2017MNRAS.471.2117A}. The MW observations were performed from October 2007 to January 2008 and a strong VHE flare was observed on December 7, 2007 by VERITAS \citep{2011ApJ...738..169A}. The measured flux was about 48\% of the Crab Nebula flux. On 10th of August 2016 the $\gamma$-ray telescope FACT detected a high emission state
which triggered the MW observations from radio to VHE $\gamma$-rays \citep{2020MNRAS.496.3912M}. 

Using the simultaneous observations of the flaring source 1ES 2344+514 from the MW, the broad-band SED was constructed and studied using both the leptonic models and the hadronic models. Recently, 
the VHE spectrum of MJD 57612 (12th August 2016) is also fitted using the one-zone SSC model, the two component SSC model and the proton-synchrotron model \citep{2020MNRAS.496.3912M,2020A&A...640A.132M}. During this observation period the VHE $\gamma$-ray flux level above 300 GeV was about 55\% of the Crab Nebula flux and the synchrotron peak frequency was shifted to $\gtrsim 10^{18}$ Hz, signalling an extreme-HBL behavior.
Although these models explain the SED well, the main difference in them lies in the strength of the magnetic field used, the energy in different components and the total number of free parameters. The one-zone SSC model employs a smaller magnetic field than the two-component model. Also, the bulk Lorentz factor in the one-zone SSC model is much higher than the two-component model. 
Low magnetic field corresponds to low magnetic field energy density leading to the electron energy density dominating over the magnetic one. The two-component model is based on the co-acceleration of protons and electrons in the jet and energy is transferred from the protons to the electrons in the shock transition layer \citep{2021A&A...654A..96Z}. The core component of the model provides seed photons for the Compton scattering, while X-rays to VHE gamma-rays are mainly produced in the blob region \citep{2020A&A...640A.132M}. Although the two-component models fit better to the observed data, they are difficult to constrain due to the large number of free parameters used.
On the other hand, the proton synchrotron model requires ultra-high energy protons and a very high magnetic field to produce VHE gamma rays. A major weakness of the proton synchrotron model is that an unrealistic super-Eddington luminosity in protons is needed to explain the observed VHE spectrum. However, in the photohadronic scenario this problem is evaded due to the higher photon density in the inner jet region \citep{Sahu:2019lwj}.

\section{photohadronic model}
We recall the salient features of the photohadronic model \citep{Sahu:2019kfd} used to explain the VHE flaring events of HBLs. As discussed earlier, when there is a temporary transition from HBL to EHBL, as in the case of Mrk 421, Mrk 501 and 1ES 1959+650, the one-zone photohadronic model is inadequate to explain the spectral behavior. In order to remedy this shortcoming the two-zone photohadronic model   is employed \citep{Sahu:2020kce,Sahu:2020tko,Sahu:2021wue} which is an extension of the one-zone photohadronic model. Depending on the nature of the VHE flaring event of the blazar, the low energy tail region of the SSC flux can either be expressed as a single power-law or a two component power-law. We refer to the former as the one-zone photohadronic model and the latter as the two-zone photohadronic model. These two models are briefly discussed below.


\subsection{One-zone photohadronic model}
The VHE flaring from a number of HBLs are successfully explained in the context of the photohadronic model. This model has also successfully explained the VHE flaring from atypical EHBLs, such as 1ES 0229+200 and 1ES 0347-232 \citep{Sahu:2019kfd}. In order to explain the VHE flaring events from HBLs, the photohadronic model assumes a double jet structure formed along the same axis. In this scenario 
a small and compact jet of size $R'_f$ is enclosed by a bigger outer jet of size
$R'_b$ with $R'_f<R'_b$. The photon densities  in the inner and the out jet are taken to be as $n'_{\gamma,f}$ and $n'_{\gamma}$ respectively where $n'_{\gamma,f}>n'_{\gamma}$ ($'$ implies in the comoving frame). The inner jet and the outer jet have bulk Lorentz factors $\Gamma_{in}$ and $\Gamma_{ext}$ respectively and for simplicity we assume 
$\Gamma_{in}\simeq \Gamma_{ext}\simeq \Gamma$ with a common Doppler factor $\mathcal{D}$ \citep{Sahu:2019lwj}. 

During the multi-TeV flaring, the Fermi accelerated high energy protons are injected into the inner jet region with a power-law spectrum $dN/dE_p\propto E^{-\alpha}_p$,
where $E_p$ is the proton energy and the spectral index $\alpha \ge 2$. These protons interact with the low energy tail region of the SSC seed photons to produce $\Delta$-resonances. The $\Delta$-resonance decays to $\gamma$-rays and neutrinos via the decay of neutral and charged pions respectively. In this decay process, each neutrino carries about half the photon energy. From the IceCube point of view, the neutrino energy is very low for the neutrinos to be detected.

The $\Delta$-resonance is produced from the $p\gamma$ interaction when the following condition is satisfied
\beq
E_{\gamma} \epsilon_\gamma=\frac{0.032\ \Gamma{\mathcal D}}{(1+z)^{2}}\ \mathrm{GeV^2}.
\label{eq:kinproton}
\eeq
Here $\epsilon_{\gamma}$ and $E_{\gamma}$ are the seed photon energy and the VHE $\gamma$-ray photon energy in the observer's frame. The quantities $\Gamma$, ${\cal D}$ and $z$ are respectively the bulk Lorentz factor, the Doppler factor and the redshift of the object. In this process about 10\% of the proton energy $E_p$ is taken by the VHE photon ($E_{\gamma}\simeq 0.1\,E_p$).
The observed VHE $\gamma$-ray flux $F_\gamma$ is proportional to the seed photon number density $n'_{\gamma,f}$ in the inner jet region 
and the incident high energy proton flux $F_p\equiv E^2_p\,dN/dE_p$.
This gives $F_\gamma\propto n'_{\gamma ,f}\,  F_{p}$.
For all the VHE flaring events observed so far from the HBLs, the range of $E_{\gamma}$ corresponds to a range of $\epsilon_{\gamma}$ that can be gotten from Eq. (\ref{eq:kinproton}), and the latter always lies in the low energy tail region of the SSC band.
The photon density in the outer jet is given by 
\beq
n'_{\gamma}(\epsilon_{\gamma})=\eta \left ( \frac{d_L}{R'_b} \right
)^2 \frac{1}{(1+z)} \frac{\Phi_{SSC}(\epsilon_{\gamma})}{{\cal
    D}^{2+\kappa}\, \epsilon_{\gamma}},
\label{photondensity}
\eeq
where  $\eta$ corresponds to the efficiency of the SSC process. We take $\eta=1$ for 100\% efficiency. $d_L$ is the
luminosity distance to the source and $\kappa=0(1)$ 
corresponds to a continuous (discrete) blazar jet.
As the inner jet expands adiabatically into the outer jet, the photon density in the former decreases. Due to uncertainties in determining the photon density in the inner region, a simple scaling law is assumed for the photon densities in the inner jet and the outer jet as follows,
\beq
\frac{n'_{\gamma,f}(\epsilon_{\gamma,1})}{n'_{\gamma,f}(\epsilon_{\gamma,2})} \simeq\frac{n'_{\gamma}(\epsilon_{\gamma,1})}{n'_{\gamma}(\epsilon_{\gamma,2})}.
\label{eq:scaling}
\eeq
This equation implies that the ratio of the photon densities for two different background energies $\epsilon_{\gamma,1}$ and $\epsilon_{\gamma,2}$ in the inner and the out jet regions are almost the same. From Eq. (\ref{photondensity}) and Eq. (\ref{eq:scaling}), $n'_{\gamma,f}$
can be expressed in terms of $\Phi_{SSC}$. This gives $F_{\gamma}\propto n'_{\gamma}$
implying that $F_{\gamma}\propto \Phi_{SSC}\epsilon_{\gamma}^{-1}$. Also, $F_\gamma \propto F_p$ implies
$F_\gamma \propto E_\gamma^{-\alpha+2}$. Thus in the low energy tail region of the SSC band the flux follows a perfect power-law behavior determined as $\Phi_{SSC}\propto \epsilon^{\beta}_{\gamma}\propto E_{\gamma}^{-\beta}$. The model based on it is referred to as the one-zone photohadronic model.  

The role of extragalactic background light (EBL) in the propagation of VHE $\gamma$-rays is crucial as it attenuates the propagating VHE photons by producing $e^+e^-$. Thus, in the observation of VHE gamma-rays, EBL effects have to be properly 
accounted for. Using different methods several EBL models have been developed \citep{1998ApJ...493..547S,Franceschini:2008tp,Dominguez:2010bv,2012MNRAS.422.3189G,2013ApJ...770...77D}. With the EBL correction, the observed VHE spectrum in the one-zone photohadronic model is given by
\beq
F_{\gamma}(E_{\gamma})=F_0 \left ( \frac{E_\gamma}{TeV} \right )^{-\delta+3}\,e^{-\tau_{\gamma\gamma}(E_\gamma,z)}=F_{\gamma, in}(E_{\gamma})\, e^{-\tau_{\gamma\gamma}(E_\gamma,z)}.
\label{eq:fluxin}
\eeq
Here $F_0$ is the normalization constant which can be adjusted from the observed VHE spectrum. In this model the only free parameter is the spectral index $\delta=\alpha+\beta$. $F_{\gamma,in}$ and $\tau_{\gamma\gamma}$ are respectively the intrinsic VHE flux and 
the optical depth for the lepton pair production process $\gamma\gamma\rightarrow e^+e^-$. For the multi-TeV flaring epochs of the HBLs, the value of $\delta$ lies in the range $2.5\le \delta \le 3.0$ \citep{Sahu:2019kfd}.

\subsection{Two-zone photohadronic model}
From recent studies on the EHBL-like behavior of the nearest blazars Mrk 421, Mrk 501 and 1ES 1959+650, it is shown that the canonical photohadronic model is not adequate to explain the VHE flaring epochs of these sources. To overcome this difficulty, it is assumed that the tail region of the SSC band which is responsible for the production of $\Delta$-resonance has two different power-laws: 
\beq
\Phi_{SSC}\propto
 \left\{ 
\begin{array}{cr}
E^{-\beta_1}_{\gamma}
, & \quad 
 \mathrm{100\, GeV\, \lesssim E_{\gamma} \lesssim E^{intd}_{\gamma}}
\\ E^{-\beta_2}_{\gamma} ,
& \quad   \mathrm{E_{\gamma}\gtrsim E^{intd}_{\gamma}}
\\
\end{array} \right. .
\label{eq:sscflux}
\eeq
The spectral indices satisfy $\beta_1\neq\beta_2$ and $E^{intd}_{\gamma}$ is the transition energy between zone-1 and zone-2. The value of $E^{intd}_{\gamma}$ is different for different spectra. Also, the value of $E^{intd}_{\gamma}$ provides information on the nature of the flaring event. Using Eq. (\ref{eq:sscflux}), we can express the observed VHE spectrum as,
\beq
F_{\gamma}=e^{-\tau_{\gamma\gamma}}\times
\begin{cases}
F_1 \, \left ( \frac{E_{\gamma}}{TeV} \right )^{-\delta_1+3}
, & \quad 
\mathrm{100\, GeV\, \lesssim E_{\gamma} \lesssim E^{intd}_{\gamma}}\,\, (\text{zone-1}) \\ 
F_2 \, \left ( \frac{E_{\gamma}}{TeV} \right )^{-\delta_2+3},
& \quad \,\,\,\,\,\,\,\,\,\,\,\, \mathrm{E_{\gamma}\gtrsim E^{intd}_{\gamma}}\,\,\, (\text{zone-2})
 \end{cases}.
\label{eq:flux}
\eeq
Thus, we now have two different zones and the normalization constants for zone-1 and zone-2 are defined as $F_1$ and $F_2$ respectively. 
As explained earlier, the photohadronic model is  well suited for $E_{\gamma} \gtrsim 100\, GeV$. Below this energy it is the leptonic model which contributes to the SED. In the context of the photohadronic model, we only concentrate on the VHE spectrum.
The observed VHE spectrum of the EHBL can be fitted in zone-1 and zone-2 separately by adjusting the spectral indices $\delta_i=\alpha+\beta_i$ ($i=1,2$) respectively. As shown, the VHE events from HBLs can be explained with $2.5\le\delta\le 3.0$ and the temporary EHBLs with the two-zone model. However, there can be another type of VHE flaring from blazars. In this situation, the spectrum can be explained by zone-2 only, i.e. 
\beq
F_{\gamma}=
e^{-\tau_{\gamma\gamma}}\times
F_2 \, \left ( \frac{E_{\gamma}}{TeV} \right )^{-\delta_2+3}.
\label{eq:flux2}
\eeq
In this case the value of the spectral index will be 
$3.1 \le \delta_2 \le 3.5$. 

We take the proton spectral index $\alpha=2$ \citep{Dermer:1993cz}. By fixing the value of $\alpha$ we automatically fix the value of $\beta_i$ for a given $\delta_i$. For the EBL correction to the VHE spectrum the EBL model of \cite{Dominguez:2010bv} and \cite{Franceschini:2008tp} are widely used. For the analysis of our results, we use the EBL model of \cite{Franceschini:2008tp}. This model is consistent with a EBL peak flux of $\sim 15\ \mathrm{nW}\mathrm{n^{-2}}\ \mathrm{sr^{-1}}$ at $1.4\, \mathrm{\mu m}$ \citep{Aleksic:2014usa,Ahnen:2016hsf} and the results are similar to the EBL model of \cite{Dominguez:2010bv}.
In earlier work \citep{Sahu:2019kfd}, we have shown that the value of $\delta_1$ (zone-1) allows one to characterize the flaring state for a given observation and is always in the range $2.5 \le \delta_1 \le 3.0$.  The standard 
VHE flaring events of all the HBLs lie in this range of $\delta_1$. To fit the data in zone-1 (low energy regime) we take this range of $\delta_1$ and fit it with different values of $F_1$. Zone-2 (high energy regime) is similarly fitted for $3.1 \le \delta_2 \le 3.5$ with different values of $F_2$. We select the best fit values of both the zones by calculating their statistical significance, the point of intersection of both the curves gives the transition energy $E^{intd}_{\gamma}$.


\begin{figure}
\centering
\includegraphics[width=.85\linewidth]{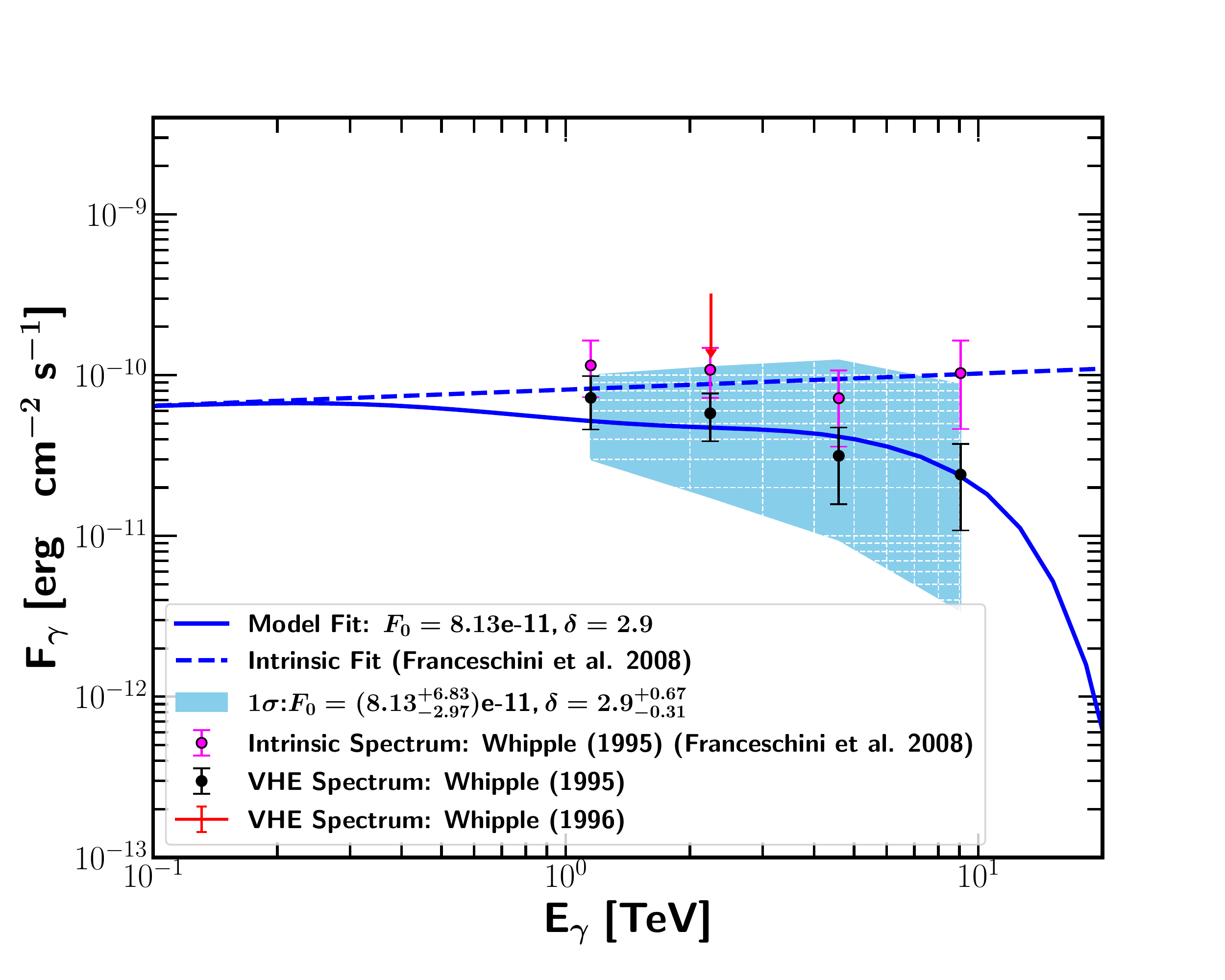}
\caption{The VHE flaring event of December 20, 1995 (MJD 50071) observed by the Whipple telescope (Wh95) \citep{Schroedter_2005} is fitted with the photohadronic model. Here and in all the subsequent figures the normalization constants $F_i$ ($i=0,2$) are expressed in units of $\mathrm{erg\, cm^{-2} \, s^{-1}}$. The best fit to the VHE spectrum is shown by the blue curve. The blue shaded region corresponds to $F_0=8.13^{+6.83}_{-2.97}\times 10^{-11}\, \mathrm{erg\, cm^{-2}\, s^{-1}}$ and $\delta=2.9^{+0.67}_{-0.31}$. The dashed curve corresponds to intrinsic flux. For reference, we have also shown the upper limit of the VHE flux for the night of 5 December 1996 observed by Whipple telescope.}
\label{fig:figure1}
\end{figure}

\section{results}
The first VHE $\gamma$-ray flaring event from the HBL 1ES 2344+514 was observed on 1995 December 20 by the Whipple 10 m telescope (Wh95)  and shown that it can be fitted with a power-law \citep{Schroedter_2005}. During this period the observed $E_{\gamma}$ was in the energy range $1.15\, TeV \lesssim E_{\gamma} \lesssim 9.1 \, TeV$ and this VHE emission was about 50 times brighter than the emission during the quiescent phase of this object measured in later years. This flux level is the historical maximum observed so far from 1ES 2344+514 and the spectrum is fitted well with $\delta=2.9$ and $F_0=8.13\times 10^{-11}\, \mathrm{erg\, cm^{-2}\, s^{-1}}$ and this corresponds to a high emission state. $F_0$ and $\delta$ are two parameters in the model and the spectral index is constrained to be in the range $2.5\le \delta \le 3.0$. By using the minimum $\chi^2$ fit and varying simultaneously both $F_0$ and $\delta$, the $1\sigma$ confidence interval defined by $\chi^2_{min}+2.3$ is calculated and the errors in these parameters are estimated for all the flaring epochs. For the above spectrum, the errors in the parameters are given by $F_0=8.13^{+6.83}_{-2.97}\times 10^{-11}\, \mathrm{erg\, cm^{-2}\, s^{-1}}$ and $\delta=2.9^{+0.67}_{-0.31}$ respectively. In Fig. \ref{fig:figure1}, the blue shaded region corresponds to the errors in the parameters $F_0$ and $\delta$. These are shown in the same figure
along with the best fit values.
In this figure we have also shown the intrinsic spectrum $F_{\gamma, in}$ (dashed curve) which is proportional to $E^{0.1}_{\gamma}$. The range $1.15\, TeV \lesssim E_{\gamma} \lesssim 9.1 \, TeV$ corresponds to the Fermi accelerated proton energy in the range $11.5\, TeV \lesssim E_p \lesssim 91 \, TeV$. $\delta=2.9$ corresponds to the flare in a high emission state and the spectral index of the background SSC photon is $\beta=0.9$. A very small magnetic field $(B\sim 10^{-4}$ G) is needed to accelerate the protons up to $\sim 100$ TeV energy in the inner jet region.

The central black hole (BH) mass of 1ES 2344+514 derived from the stellar velocity dispersion measurements is $M_{BH}\sim 10^{8.80\pm0.16}\, M_{\odot}$ \citep{2003ApJ...583..134B} which corresponds to an Eddington luminosity $L_{Edd}\sim (0.55 - 1.15)\times 10^{47}\, \mathrm{erg\, s^{-1}}$. The integrated flux in the energy range $1.15\, TeV \lesssim E_{\gamma} \lesssim 9.1 \, TeV$ is $F_{\gamma} \sim 8.53\times 10^{-11} \mathrm{erg\, cm^{-2}\, s^{-1}}$, corresponding to a luminosity of $L_{\gamma}\sim 4.1\times 10^{44}\,\mathrm{erg\, s^{-1}}$.

We can estimate the bulk Lorentz factor $\Gamma$ for this flaring event. As shown in Fig. 7 of \cite{2011ApJ...738..169A}, the tail region of the SSC starts around $10^{20}$ Hz. By assuming that the 9.1 TeV observed  $\gamma$-ray is produced from the interaction of 91 TeV proton with the SSC seed photons of energy $\epsilon_{\gamma} \gtrsim 2\times 10^{20}$ Hz ($\gtrsim 0.83$ MeV), we obtain the bulk Lorentz factor $\Gamma \gtrsim 16$ which is consistent with the X-ray variability time scale of about an  hour as seen from 1ES 2344+514 \citep{2000MNRAS.317..743G}. 

To produce the $\Delta$-resonance within the jet, we take $R'_f\sim 5\times 10^{15}\, \mathrm{cm}$ and $R'_b\sim 10^{16}\, \mathrm{cm}$ and the optical depth $\tau_{p\gamma}$ for the $p\gamma\rightarrow\Delta^+$ process should satisfy $\tau_{p\gamma} < 1$. Also the proton luminosity $L_p=7.5\tau_{p\gamma}^{-1} L_{\gamma}$ should satisfy $L_p < L_{Edd}/2$. By taking $L_{Edd}\sim  10^{47}\, \mathrm{erg\, s^{-1}}$ \citep{2003ApJ...583..134B}, we obtain $\tau_{p\gamma} > 0.06$. By taking $\tau_{p\gamma}\sim 0.2$, the SSC photon density in the inner jet region is 
$n^{'}_{\gamma,f}\sim 8.0\times 10^{10}\, \mathrm{cm^{-3}}$ and  $L_p\sim 1.5\times 10^{46}\, \mathrm{erg\, s^{-1}}$. Similarly, we can estimate the photon density and the $L_p$ for other flaring epochs also.

In the year 1996 the X-ray spectrum observed was in the EHBL range (especially the synchrotron peak) and also during this period the Whipple telescope provided the upper limit to VHE spectrum which we show in Fig. \ref{fig:figure1} for comparison. In Fig. 7 of \cite{2007ApJ...662..892A}, the Wh95 spectrum is fitted with the one-zone SSC model. However, we observe that this spectrum is not consistent with the original spectrum \citep{Schroedter_2005}. Therefore, we do not compare our results with the one-zone leptonic model fit of \cite{2007ApJ...662..892A}.

\begin{figure}
\centering
  \includegraphics[width=.85\linewidth]{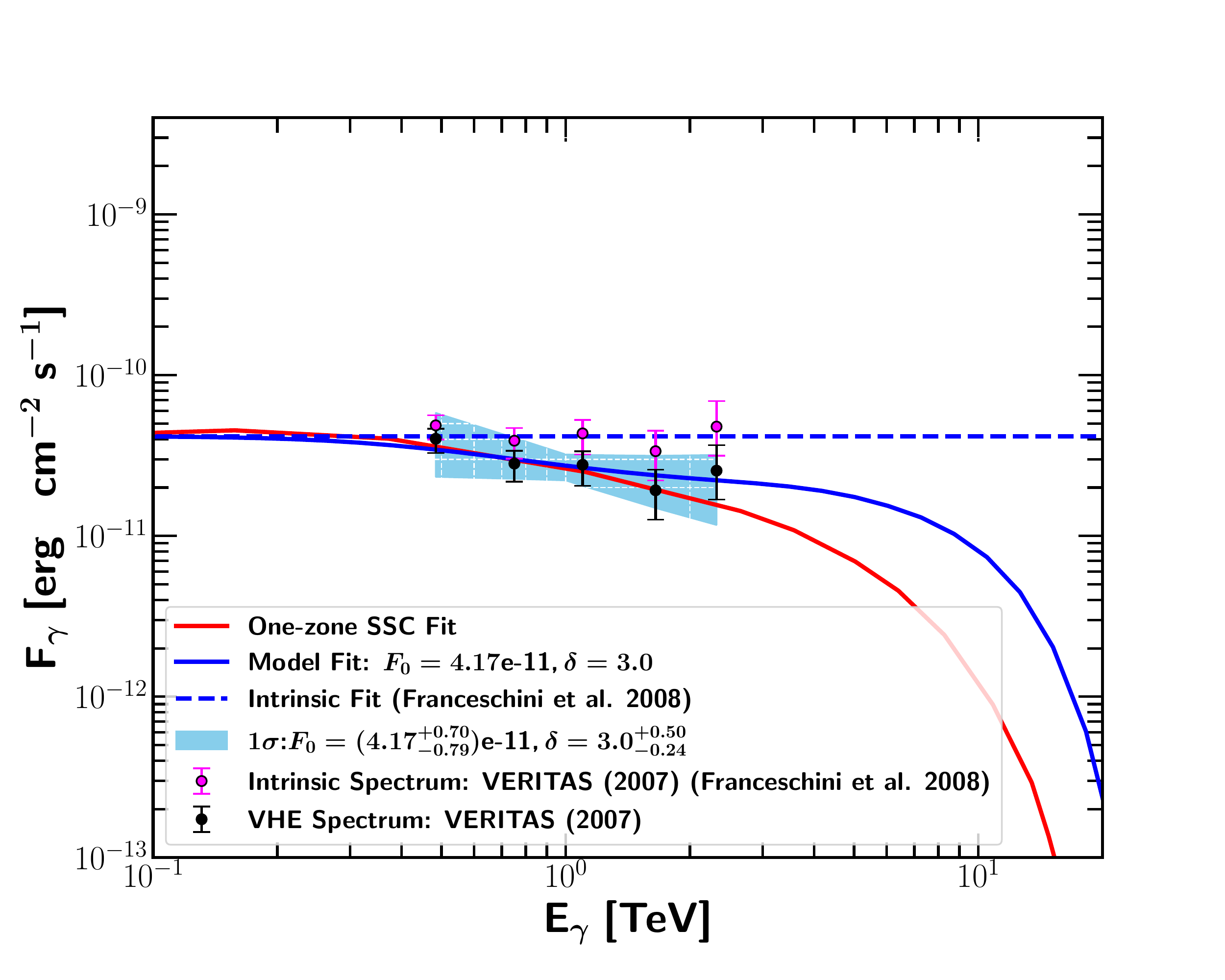}
\caption{The VHE flaring event of 7th December 2007 (MJD 54441) observed by VERITAS telescopes is fitted with the photohadronic model. The blue shaded region corresponding to the errors in the parameters $F_0$ and $\delta$ are shown at $1\sigma$ confidence interval. The dashed curve corresponds to intrinsic flux. For comparison, we have also shown the one-zone SSC fit \citep{2011ApJ...738..169A}
}
\label{fig:figure2}
\end{figure}

The VERITAS telescopes observed 1ES 2344+514 for 37 nights during a multiwavelength observation period between 4th October 2007 and 11th January 2008. On 7th of December 2007, VERITAS observed a strong VHE flare and the flux above 300 GeV was about 48\% that of the Crab Nebula flux. This VHE flare was observed in the energy range 
$0.48\, TeV \lesssim E_{\gamma} \lesssim 2.3 \, TeV$. Also, during this period, the synchrotron peak was in the HBL range. 
In the context of the photohadronic model, the VHE spectrum is fitted very well with the spectral index $\delta=3.0$ and the normalization factor $F_0=4.17\times 10^{-11}\mathrm{erg\, cm^{-2}\, s^{-1}}$.
The errors in the parameters are given by $F_0=4.17^{+0.70}_{-0.79}\times 10^{-11}\, \mathrm{erg\, cm^{-2}\, s^{-1}}$ and $\delta=3.0^{+0.50}_{-0.24}$ respectively. In Fig. \ref{fig:figure2}, we present the fit to the spectrum, the intrinsic spectrum and the blue shaded region in the figure
corresponds to the $1\sigma$ errors in $F_0$ and $\delta$. The spectrum is also fitted with the one-zone SSC model \cite{2011ApJ...738..169A}. It is observed that our fit is marginally better than the one-zone SSC fit as shown in Fig. \ref{fig:figure2}. For $\Gamma={\cal D}=16$, we obtain $3.3\, MeV \lesssim \epsilon_{\gamma}\lesssim 15.7\, MeV$.

Also, the fitted spectrum in the one-zone SSC model falls faster than the photohadronic fit for $E_{\gamma} \gtrsim 2.3$ TeV. It was observed that by excluding the flux of the flaring event, the average flux during this observation period was found to be low. In the photohadronic model, the integrated flux and the corresponding luminosity in the energy range $0.48\, TeV \lesssim E_{\gamma} \lesssim 2.3 \, TeV$ are respectively  $F_{\gamma} \sim 4.28\times 10^{-11} \mathrm{erg\, cm^{-2}\, s^{-1}}$ and $L_{\gamma}\sim 2.05\times 10^{44}\,\mathrm{erg\, s^{-1}}$.

\begin{figure}
\centering
  \includegraphics[width=.85\linewidth]{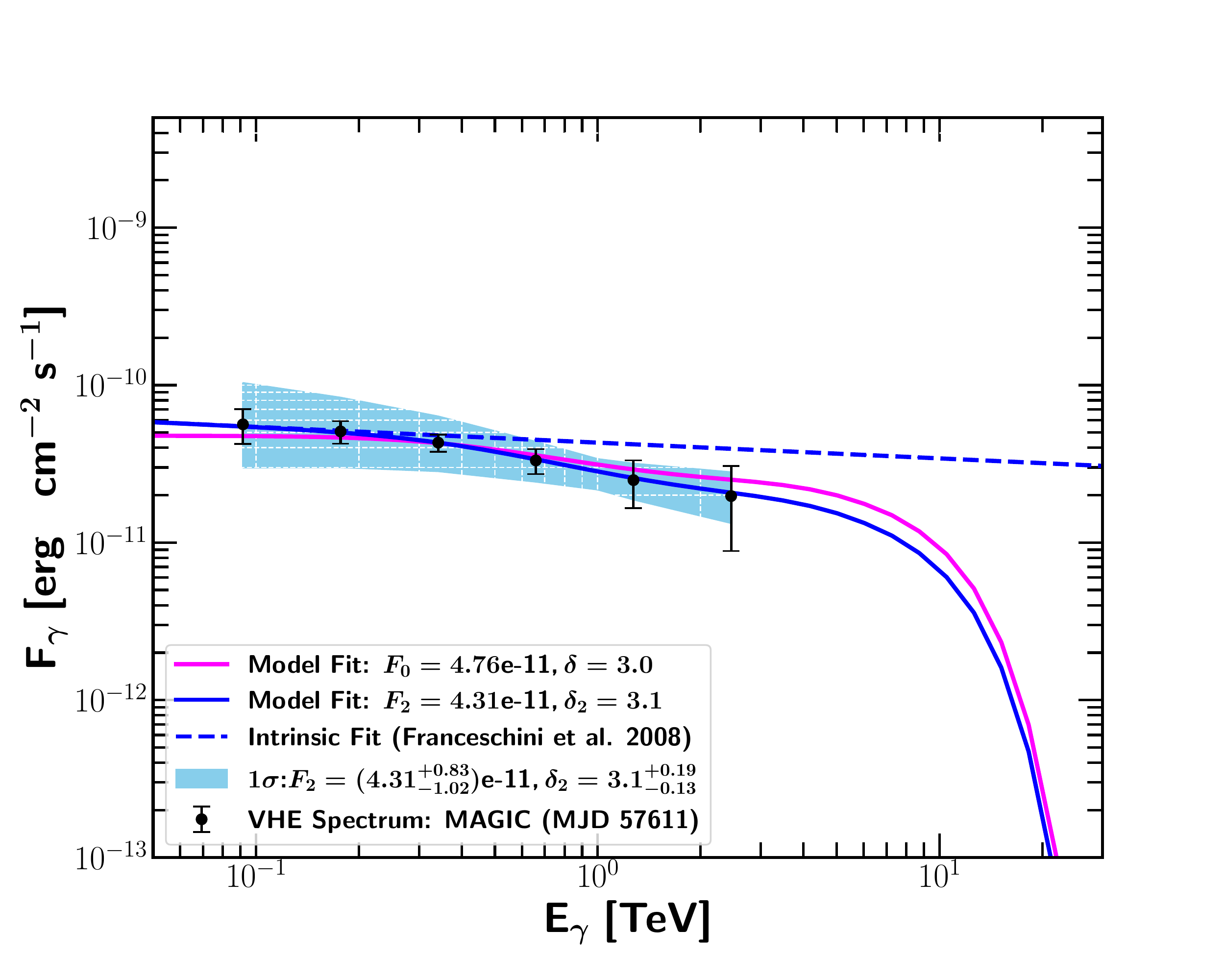}
\caption{The VHE spectrum of MJD 57611 (11th August 2016) observed by MAGIC is fitted with the photohadronic model ($\delta=3.0$ and $F_0=4.76\times 10^{-11}\, \mathrm{erg\, cm^{-2}\, s^{-1}}$) and with the two-zone photohadronic model ($\delta_2=3.1$ and $F_2=4.31\times 10^{-11}\, \mathrm{erg\, cm^{-2}\, s^{-1}}$ corresponds to zone-2). Here zone-1 is absent. The dashed curve is the intrinsic spectrum corresponding to two-zone model. We have also shown in the blue shaded region the errors in $F_2$ and $\delta_2$ at the $1\sigma$ confidence interval.}
\label{fig:figure3}
\end{figure}

The EHBL-like flaring event of 10th August 2016 was first observed in VHE by FACT and follow-up observations were undertaken in MW by various telescopes. Particularly in VHE it was observed by the MAGIC telescopes on MJD 57611 and MJD 57612 (11th and 12th of August 2016). For MJD 57611 the source was observed for 0.39 hours in the energy range $0.09\, TeV \lesssim E_{\gamma} \lesssim 2.5 \, TeV$ and 
high flux was recorded. On MJD 57612, the source was observed for  0.48 hours and the observed integral VHE flux above 300 GeV was $(2.1\pm 0.4)\times 10^{-11}\mathrm{cm^{-2}\, s^{-1}}$ which is about three times lower than the flux observed during the previous night. 
The synchrotron peak frequency observed during this period was above $10^{18}$ Hz and according to the classification scheme this belongs to the EHBL category. 
 The flaring events of MJD 57611 and MJD 57612 have similar physical conditions to that of EHBL-like VHE flaring events of Mrk 421, Mrk 501 and 1ES 1959+650 and the latter events are explained with the two-zone photohadronic scenario \citep{Sahu:2020tko,Sahu:2020kce,Sahu:2021wue}. Thus it is imperative to use again the two-zone photohadronic model to interpret the VHE spectra of MJD 57611 and MJD 57612 where the zone-1 has spectral index $\delta_1$ and the zone-2 has spectral index $\delta_2$. Both these zones have
the common transition energy $E^{intd}_{\gamma}$. We first tried to fit the spectrum of MJD 57611 with the two-zone photohadronic model as discussed above by using Eq. (\ref{eq:flux}). However, it is observed that the whole spectrum can be fitted perfectly well only with the parameters corresponding to zone-2, i.e., $\delta_2=3.1$ and its corresponding normalization constant $F_2=4.31\times 10^{-11}\mathrm{erg\, cm^{-2}\, s^{-1}}$ which is shown in Fig. \ref{fig:figure3}. Thus, in this case zone-1 appears to be depleted and therefore there is no need for $E^{intd}_{\gamma}$. 

Recently, from the EHBL-like VHE flaring of Mrk 421 \citep{Sahu:2021wue} which was observed for 13 consecutive days, we have shown that the VHE flaring event decreases from a high emission state to a low emission state, and correspondingly, the transition energy $E^{intd}_{\gamma}$ starts with $\sim 1$ TeV and slowly decreases to $\sim0.25$ TeV. Consequently, during this transition process, zone-2 became wider and spread into zone-1, making the latter region narrower at the end of the flaring process. 

It was expected that after the first observation of the
EHBL-like behavior on 10th of August 2016 from 1ES 2344+514,
the VHE spectrum observed in the subsequent two days (11th and 12th) should be fitted
with two different zones with a transition energy $E^{intd}_{\gamma}$
and in the subsequent days $E^{intd}_{\gamma}$ should have moved towards lower energy.
However, unexpectedly in the case of MJD 57611, $E^{intd}_{\gamma}$
does not exist and zone-1 is absent in the spectrum.
Thus, for the spectrum of MJD 57611, it is possible that the zone-1 became too narrow or was completely taken over by zone-2 in one day, making the spectrum fit well with the spectral index of zone-2 only. 

We have also fitted the spectrum using the one-zone photohadronic model and good fit is obtained for $\delta = 3.0$ and $F_0=4.76\times 10^{-11}\, \mathrm{erg\, cm^{-2}\, s^{-1}}$ which is consistent with a low emission state. Even though the spectrum seems to be compatible with the low emission state interpretation (with $\delta=3.0$) of a HBL, the physical conditions of the flaring event are those of the EHBL type. Thus, the standard HBL interpretation is discarded. For comparison, we have also shown this fit in Fig. \ref{fig:figure3}. The blue shaded region in this figure corresponds to the errors in parameters $\delta_2=3.1^{+0.19}_{-0.13}$ and $F_2=4.31^{+0.82}_{-1.02}\times 10^{-11}\, \mathrm{erg\, cm^{-2}\, s^{-1}}$ at the $1\sigma$ confidence level. The integrated flux on MJD57611 and its corresponding luminosity are  respectively, $F_{\gamma}=1.25\times 10^{-10}\mathrm{erg\, cm^{-2}\, s^{-1}}$ and $L_{\gamma}\sim 6.0\times 10^{44}\,\mathrm{erg\, s^{-1}}$.

\begin{figure}
\centering
  \includegraphics[width=.85\linewidth]{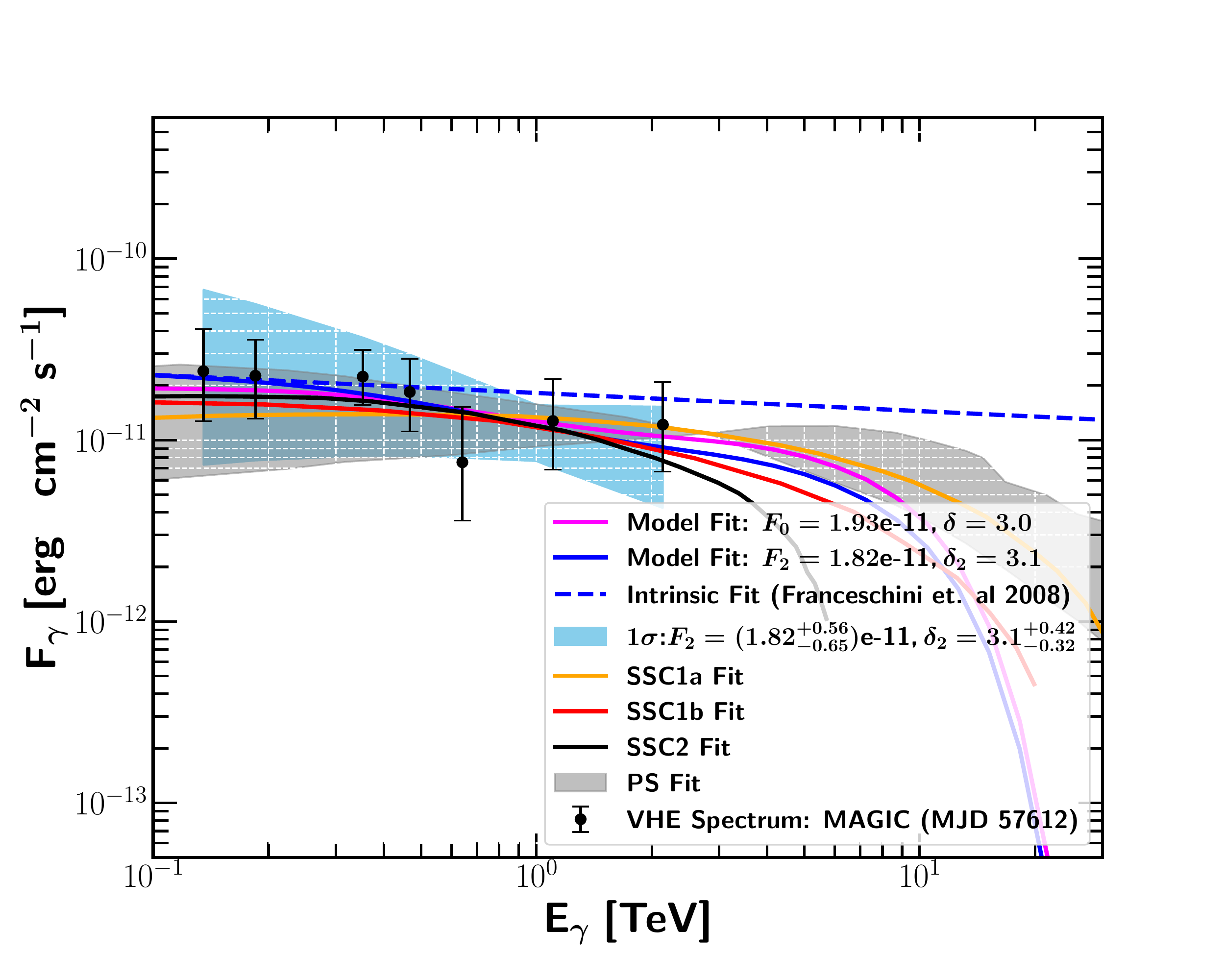}
\caption{The VHE spectrum of MJD 57612 (12th August 2016) is fitted with the one-zone ($\delta=3.0$ and $F_0=1.93\times 10^{-11}\, \mathrm{erg\, cm^{-2}\, s^{-1}}$) and the two-zone ($\delta_2=3.1$ and $F_2=1.82\times 10^{-11}\, \mathrm{erg\, cm^{-2}\, s^{-1}}$) photohadronic model. 
The dashed curve is the intrinsic spectrum corresponding to the two-zone photohadronic model. The blue shaded region in this figure corresponds to the errors in parameters $\delta_2=3.1^{+0.42}_{-0.32}$ and $F_2=1.82^{+0.56}_{-0.65}\times 10^{-11}\, \mathrm{erg\, cm^{-2}\, s^{-1}}$ at the $1\sigma$ confidence level. We compare the photohadronic fit with the one-zone SSC model (SSC1a and SSC1b), the two component SSC model (SSC2) and the proton-synchrotron model (PS) of \cite{2020MNRAS.496.3912M} and \cite{2020A&A...640A.132M}. The proton synchrotron model obtained by varying the maximum proton-synchrotron peak frequency, radius of the emitting region and the proton normalization is represented by the gray shaded region.
}
\label{fig:figure4}
\end{figure}

\begin{table}
\centering
\caption{In Fig. \ref{fig:figure4} the VHE spectrum of MJD 57612 is fitted with various models: the two different one-zone SSC model (SSC1a and SSC1b), the two-component SSC model (SSC2) (blob+core), the proton synchrotron model (PS) \citep{2020MNRAS.496.3912M,2020A&A...640A.132M} and the photohadronic model (PH). The various parameters, like the bulk Lorentz factor ($\Gamma$), the spectral index ($\delta$), the blob Radius ($R'_b$ in units of $10^{16}$ cm ) and the magnetic field ($B'$ in G) used in these models are summarized below. The value of the spectral index $\delta$ given here is for the photohadronic model for MJD 57611 and MJD 57612 when the object was behaving as EHBL.
}
\begin{tabular*}{\columnwidth}{@{\extracolsep{\fill}}llllll@{}}
\hline
\multicolumn{1}{@{}l}{Parameter} & SSC1a & SSC1b & SSC2 & PS & PH\\
  & & & blob, core & &\\
\hline
\hline
$\Gamma$ & 30 & 20 & 6, 4 & 30 & $\gtrsim 16$\\
$\delta$ & 2  & 2.07 & --,-- & 1.5-2.5  & 3.1\\
$R'_b$ & 1 & 1.22 & 1.07, 16.0 & 0.09-0.19 & 1\\ 
$B'$ & 0.02 & 0.02 &0.1, 0.1 & 48-63 & $\sim 10^{-4}$\\ 
\hline
\end{tabular*}
\label{tab:tab1}
\end{table}

The observed VHE spectrum of MJD 57612 was in the energy range $0.14\, TeV \lesssim E_{\gamma} \lesssim 2.1 \, TeV$ and it has exactly the same behavior as MJD 57611. 
The observed spectrum on this night is fitted very well by zone-2 parameters $\delta_2=3.1$ and $F_2=1.82\times 10^{-11}\mathrm{erg\, cm^{-2}\, s^{-1}}$. The errors in these parameters at $1\sigma$ confidence level are given by
$\delta_2=3.1^{+0.42}_{-0.32}$ and $F_2=4.31^{+0.56}_{-0.65}\times 10^{-11}\, \mathrm{erg\, cm^{-2}\, s^{-1}}$. The best fit to the spectrum and the blue shaded region corresponding to the errors in $F_2$ and $\delta_2$ are shown in Fig. \ref{fig:figure4} along with the intrinsic spectrum. 
 For comparison, we have shown the one-zone photohadronic fit with $\delta=3.0$ and $F_0=1.93\times 10^{-11}\mathrm{erg\, cm^{-2}\, s^{-1}}$ which fits well to the observed spectrum as shown in Fig. \ref{fig:figure4}. However, in this case also, we have discarded this fit for the same reason given for the spectrum of previous night (MJD 57611).
 
In previous studies, the VHE spectrum of MJD 57612 was fitted using the one-zone SSC model, the two-component SSC model (SSC2) and the proton-synchrotron model (PS) \citep{2020MNRAS.496.3912M,2020A&A...640A.132M}. In the one-zone SSC model, two different sets of parameters were used to fit the spectrum. For convenience, we make use of the following labels, SSC1a \citep{2020MNRAS.496.3912M} and SSC1b \citep{2020A&A...640A.132M} as reference labels. Although, the overall fit to the spectrum is good in all these models as shown Fig. \ref{fig:figure4} , they behave differently in the high energy regime and also the parameters used are different. In Table \ref{tab:tab1}, we have summarized the parameters used in these models. For $E_{\gamma}\gtrsim 1$ TeV all models start behaving differently as shown in Fig. \ref{fig:figure4}. The SSC2 fit falls faster than the rest of the fits \citep{2020A&A...640A.132M}. 
The proton-synchrotron model of \cite{Cerruti:2014iwa} is employed to fit the VHE spectrum. By varying the maximum proton-synchrotron peak frequency, radius of the emitting region and the proton normalization, the shaded region is constructed in Fig. \ref{fig:figure4}. Although the leptonic and the proton-synchrotron models fit the data well, the values of the magnetic field used are different and the difference commences for $E_{\gamma}\gtrsim 2$ TeV. The photohadronic fit lies between the SSC2 fit and the SSC1a fit \citep{2020MNRAS.496.3912M}. Also, for this night the integrated flux and the luminosity are respectively, $F_{\gamma}=4.21\times 10^{-11}\mathrm{erg\, cm^{-2}\, s^{-1}}$ and $L_{\gamma}\sim 2.02\times 10^{44}\,\mathrm{erg\, s^{-1}}$. 

In the context of the photohadronic scenario, the characteristic feature of the two flaring events (MJD 57611 and MJD 57612) is that, during the temporary EHBL-like behavior of the source, their intrinsic spectra are proportional to $E^{-0.1}_{\gamma}$ and the seed photon flux in the SSC region follows $\Phi_{SSC}\propto \epsilon^{1.1}_{\gamma}$. The spectra of both the nights are fitted very well with the parameters from zone-2 ($\delta_2$ and $F_2$) only. 
It is important to mention that, had the source not been observed in the X-ray region to see the synchrotron peak above $10^{18}$ Hz, it would have easily mislead one to interpret the VHE flaring events as standard HBL flaring events as both these spectra can be explained well with the photohadronic model for $\delta=3.0$, corresponding to low emission state (as shown in Figs. \ref{fig:figure3} and \ref{fig:figure4}). Previously, in the context of Mrk 421, Mrk 501 and 1ES 1959+650, we have shown that the VHE spectrum from temporary EHBL-like behavior can be explained by the two-zone photohadronic model where the spectral indices $\delta_1$ and $\delta_2$ contribute to the fit of the spectrum \citep{Sahu:2020tko,Sahu:2020kce,Sahu:2021wue}. However, to our knowledge, fitting the VHE spectrum of a temporary EHBL with the spectral index $\delta_2$ only, so far has not been observed in any of the EHBL-like events identified in other nearby objects. Probably this is an indication of a new type of temporary EHBL whose VHE spectrum is explained by a single zone-2 spectral index with $\delta \gtrsim 3.1$.

In the photohadronic model, neutrino energy expectations are $E_{\nu}=0.5\,E_{\gamma}$. On December 20, 1995, the HBL 1ES 2344+514 attained the highest flux level and during this flaring period, the maximum $\gamma$-ray energy observed was $E_{\gamma}\simeq 9.1 \, TeV$. This corresponds to $E_{\nu}\simeq 4.5 \, TeV$ which is very low for neutrinos to be observed in a neutrino detector like IceCube. Furthermore, the  luminosity distance of the HBL 1ES 2344+514 is $d_L \simeq 200 \, Mpc$ which will further deplete the neutrino flux.  During the flaring events of December 2007 and August 2016, the maximum observed $E_{\gamma}$ was much lower than $9.1 \, TeV$ which correspond to $E_{\nu} \ll 4.5 \, TeV$. This implies that it is difficult to observe high energy neutrinos from the above VHE flaring events.

\section{Conclusion}
In conclusion, the HBL 1ES 2344+514 was first detected in 1995 and since then several episodes of VHE flaring have been reported. In 1996 a flare in the X-ray was observed and the synchrotron peak of the SED was found to be in the EHBL category. Similarly, an outburst in the VHE was observed by FACT during August 2016 and once again the emission was EHBL-like. 

Using the photohadronic model we have attempted to explain the VHE emissions of the 20th of December 1995 observed by Whipple and the 7th of December 2007 observed by VERITAS respectively. As the VHE spectra of 
MJD 57611 and MJD 57612 are EHBL-like, we have used the two-zone photohadronic model to explain these spectra. However, we observe that the low energy regime i.e. zone-1 of these spectra are absent in these fittings and only zone-2 with $\delta_2=3.1$ is relevant to fit the observed spectra. The photohadronic fit to the VHE spectrum of MJD 57612 is also compared with the leptonic and the proton-synchrotron models.
To our knowledge, the VHE flaring events of MJD 57611 and MJD 57612 seem like new type of events originating from a transient EHBL-like source and their observed VHE spectra can be deduced from Eq. (\ref{eq:fluxin}) by taking 
$\delta\gtrsim 3.1$. So far
this type of behavior has not been observed in any of the EHBL--like events from nearby objects. We believe that these events are of the new type. However, in order to strengthen our claim, more data on simultaneous observations on other potential sources are needed in the future.

\section*{Acknowledgements}
\addcontentsline{toc}{section}{Acknowledgements}
We thank the referee for suggetions and constructive remarks to considerably improve the manuscript. The work of S.S. is partially supported by DGAPA-UNAM (Mexico) Project No. IN103522. Partial support from CSU-Long Beach is gratefully acknowledged.

\section*{Availability of data}
No new data were generated or analysed in support of this research.
\bibliography{ref}{}
\bibliographystyle{aasjournal}
\end{document}